\documentclass{iaus}
\usepackage{graphicx}

\title[Circumstellar Nebulae of Evolved Massive Stars] 
{Infrared Tracers of Mass-Loss Histories and Wind-ISM Interactions in Hot Star Nebulae}

\author[P. Morris \& the {\em{Spitzer}} WRRINGS team]   
{Patrick Morris$^1$ \and the {\em{Spitzer}} WRRINGS team}

\affiliation{$^1$NASA Herschel Science Center, IPAC, Caltech, M/C 100-22, Pasadena, CA 91125
\\ email: {\tt pmorris@ipac.caltech.edu} \\[\affilskip]
}

\pubyear{2008}
\volume{250}  
\pagerange{1--6}
\jname{Massive Stars as Cosmic Engines}
\editors{F. Bresolin, P. Crowther \& J. Puls, eds.}
\begin{document}

\maketitle

\begin{abstract}
Infrared observations of hot massive stars and their environments provide a detailed picture of mass
loss histories, dust formation, and dynamical interactions with the local stellar medium that can 
be unique to the thermal regime.  We have acquired new infrared spectroscopy and imaging  
with the sensitive instruments onboard the {\em{Spitzer}} Space Telescope in guaranteed and open time 
programs comprised of some of the best known examples of hot stars with circumstellar nebulae, 
supplementing with unpublished Infrared Space Observatory spectroscopy.  Here we
present highlights of our work on the environment around the extreme P Cygni-type star 
HDE316285, providing some defining characteristics of the star's evolution and interactions 
with the ISM at unprecented detail in the infrared.

\end{abstract}

\firstsection 
\section{Introduction}

Observations of the circumstellar environments of Lumnious Blue Variables (LBVs) with ESA's
Infrared Space Observatory (ISO) and modern ground-based imaging devices have demonstrated 
how sensitive mid-infrared imaging and spectroscopy can substantially improve our knowledge 
of the distribution of material, the physical and chemical properties of nebular dust and gas, 
and therefore on the history of mass loss from the surface of the massive central star during
its evolution from the Main Sequence.  Other than the bright and compact environments of most
LBV nebulae, none of the lower surface brightness nebulae around OB or Wolf-Rayet stars 
could be observed with ISO due mainly to instrumental sensitivity limitations. 
The instruments on the {\em{Spitzer}} Space Telescope have since offered significant gains in
sensitivity and sky coverage; for comparison, the {\em{Spitzer}} Infrared Spectrometer (IRS;
5.3 -- 38 $\mu$m) is factor of $\sim$50-100 more sensitive than the ISO Short Wavelength 
Spectrograph (SWS; 2.4-45.2 $\mu$m) at 5 $\mu$m, depending on resolution modes of the two instruments.  
We have exploited the {\em{Spitzer}} IRS,  the Infrared
Array Camera (IRAC) with imaging bands at 3.4, 4.6, 5.8, and 8.0 $\mu$m, and the Multiband Imaging
Photometer for {\em{Spitzer}} (MIPS) with imaging capabilities at 24, 70, and 160 $\mu$m in
Guaranteed and Open Time programs to observe the environments around a number of hot, massive
stars.  Most of these stars are surrounded by ring nebulae and are well known for their optical 
and/or radio properties; NGC2359 (WR), M1-67 (WR), G79.29+00.46 (BIe), and HD148937 (Ofp?e) are
examples.  

In this paper we summarize our study of the massive B supergiant and candidate LBV 
HDE316285 from our program.   This highly luminous P Cyg-type star has been suspected of
being surrounded by an extended, cold nebula (\cite[McGregor, Hyland, \& Hillier 
1998]{McGregorHylandHillier_98}); we have obtained conclusive observations of the environment
around this star, which harbor telling characteristics of the star's mass loss history, wind-wind
and wind-ISM interactions, the molecular content of the gas and thermal history of the dust. 

\begin{figure}[b]
\begin{center}
 \includegraphics[width=3.4in]{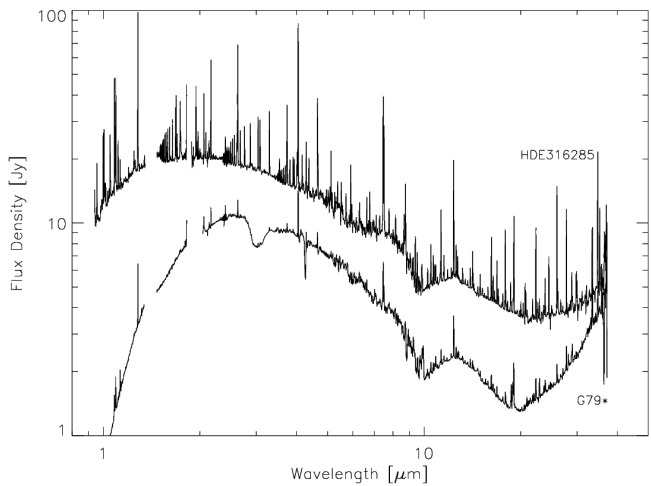} 
 \caption{Combined near-IR (1.4-2.4$\mu$m), ISO/SWS (2.4-10$\mu$m), and {\em{Spitzer}}/IRS 
 (10.0-38$\mu$m) SEDs of HDE316285 (upper) and G79.29+00.46 (lower). From Morris et al. (2008) 
 submitted.}
   \label{fig1}
\end{center}
\end{figure}

\section{The ring nebula around the extreme P Cygni star HDE316285}

The combined 1.4-38 $\mu$m spectrum of HDE316285 is shown in Fig. \ref{fig1}.  This star has been
quantitatively characterized by \cite[Hiller et al. (1998)]{Hillier_etal98} using line-blanketed
non-LTE wind models as an ``extreme'' 
P Cygni-type star, with an optical and near-IR stellar wind spectrum that is quite similar
to that of P Cyg.  Spectral variability, the stellar 
properties and chemical content of HDE 316285 are similar to known LBVs, but is more extreme 
than P Cyg because of its high wind performance number (= ratio of wind momentum to radiative 
momentum), some 30 times 
greater.   This number, however, is based on the assumption of a smooth, homogeneous wind and
is subject to downward revision in the case of a clumped wind and associated reduction to the
mass loss rate (cf. \cite[Hillier\&Miller 1999]{HillierMiller99}). Somewhat unexpectedly, 
however, the infrared continuum, which can be used to constrain the volume filling factor (e.g., 
\cite[Morris et al. 2000]{Morris_etal00}), 
is contaminated by a strong thermal excess which becomes evident at around 
16$\mu$m (see Fig. \ref{fig1}).
In fact \cite[McGregor, Hyland, \& Hillier (1988)]{McGregorHylandHillier88} had pointed out 
that HDE316285 is a moderate IRAS source, exhibiting cold dust ($\sim$60 K) removed from the 
central star.  Since this star is projected $\sim1^\circ$ from the Galactic Center and is now recognized 
to be in close (apparent) proximity 
to the infrared star forming region (SFR) Sgr~D (discussed below), a significant thermal component
arising from surround molecular material heated by nearby young stars would provide the 
most likely explanation for the excess emission in the
IRAS 25 and 60 $\mu$m passbands.

\begin{figure}[t]
\begin{center}
 \includegraphics[width=1.75in]{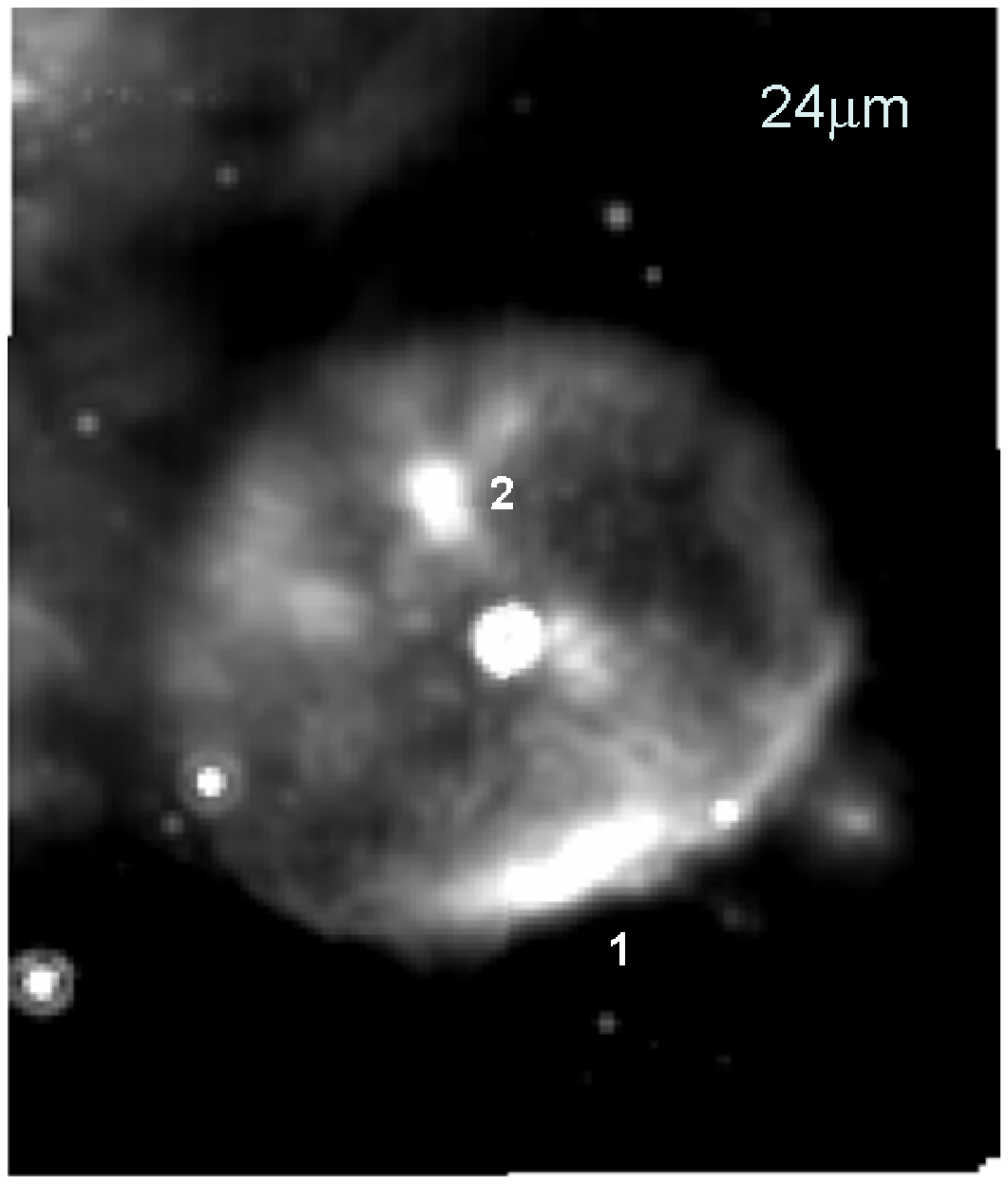}\includegraphics[width=2.7in]{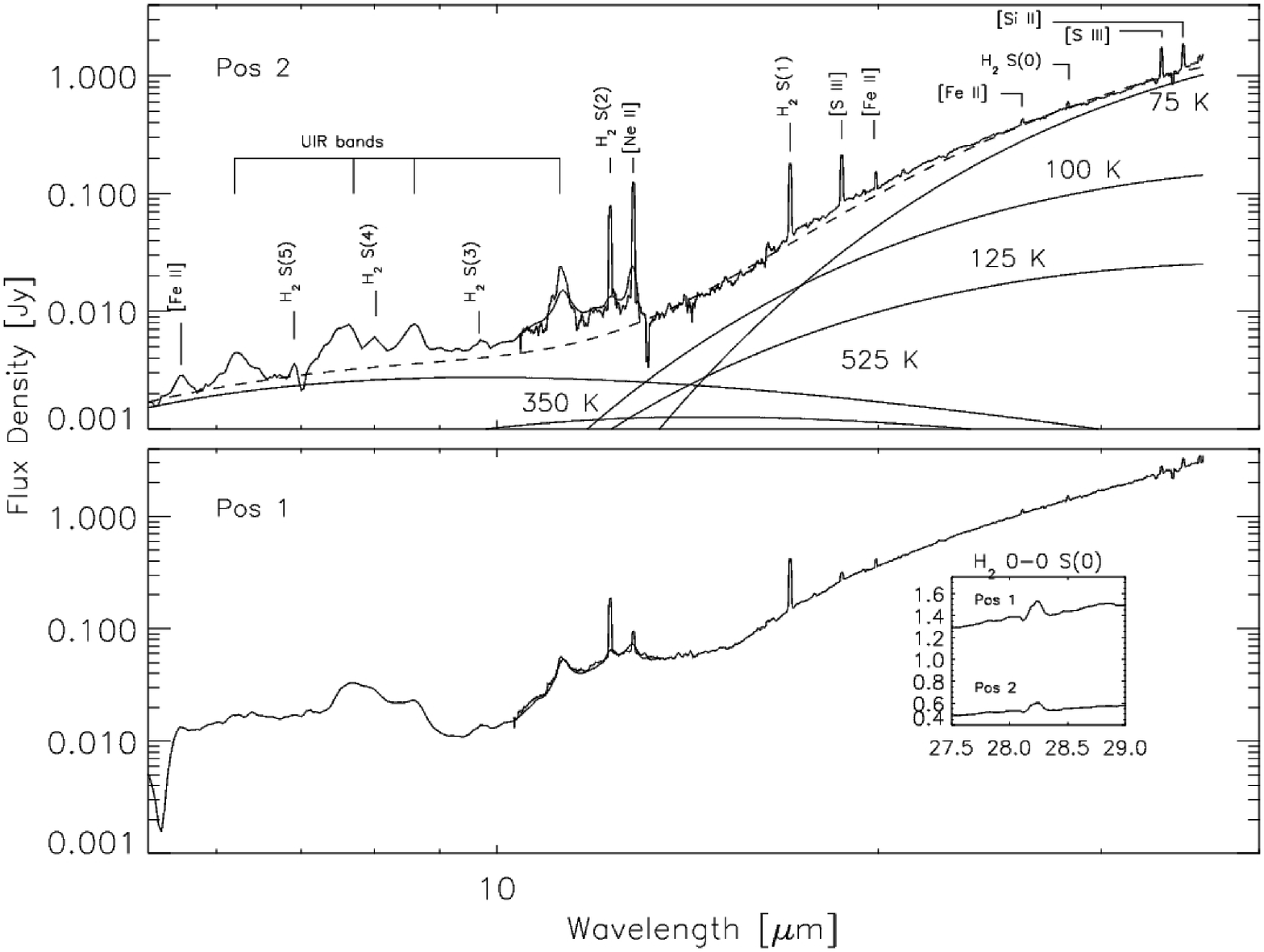}  
 \caption{{\em{Left:}} HDE316285 at 24 $\mu$m with Spitzer/MIPS. The image is $5'.5 \times 8$, north 
is up, east is left.  Two regions targeted for spectroscopy are indicated. 
{\em{Right:}} IRS 5.3-38$\mu$m spectra at two positions in the nebula (background-subtracted), 
with main features identified.}
   \label{fig2}
\end{center}
\end{figure}

\begin{figure}[t]
\begin{center}
\includegraphics[width=3.8in]{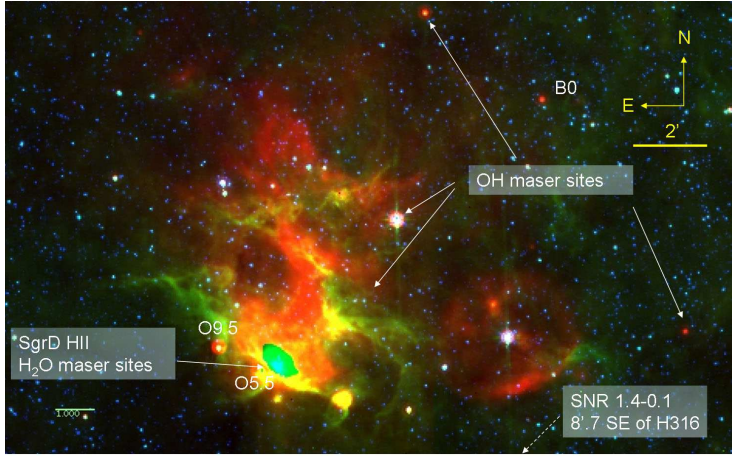}
\caption{Mosaic of the Sgr~D region, with radio sources identified by 
 \cite[Mehringer et al. (1998)]{Mehringer_etal98} indicated.  Blue corresponds to 4.6 $\mu$m, 
 green to 8.0 $\mu$m, and red to 24 $\mu$m.}
  \label{fig3}
\end{center}
\end{figure}

{\underline{\it{Infrared imaging}}}.  We now confirm that HDE316285 is definitely surrounded 
by its own circumstellar nebula. 
Following careful inspection of archival Midcourse Space Experiment 21.3 $\mu$m data in which 
a bipolar nebula can be seen just barely above the detection threshhold, 
our MIPS 24$\mu$m imaging (see Fig. \ref{fig2}) reveals a limb-brightened, 
3$'.4 \times 4'.1$ ring nebula with the major axis oriented $\sim40^\circ$ from the north.  
The nebula is {\em{not}} discernable overall 
in the IRAC bands, though one bright knot of 8.0 $\mu$m emission is detected 
$\sim1'$ NE of the central star. This knot and the brightened southern limb (respectively labelled 
''2'' and ''1'' in Fig. \ref{fig1}) exhibit peak nebular fluxes of 624 and
450 MJy/sr, and were targeted by us for IRS spectroscopy, shown in Fig. \ref{fig2} and discussed 
below.  The nebula at 24 $\mu$m also clearly exhibits a clumpy structure, and filaments directed 
radially outwards particularly evident in the NE half.  The SW rim of the nebula appears to
be interacting with material emitting at 8.0 $\mu$m in the surrounding ISM. (see also Fig. \ref{fig3}).

{\underline{\it{Spectral content: PAHs and rotational H$_2$.}}}
Spectra extracted at the two positions indicated in Fig. \ref{fig2} and fully corrected for
background emission with a dedicated off-position observation reveal thermal spectra with two
principle components: a dominant cool component in the 75-125 K range, and a warm component at
$T_{\rm{dust}} >$ 350 K.  Atop the warm component we see the family of polyaromatic hydrocarbon
(PAH) bands at 6.22, 7.63, 8.63, 11.22, and 12.75 $\mu$m, and possibly a number of secondary peaks.
These bands are common in reflection nebulae and the shells of AGB and post-AGB stars with mixed
C and O chemistries, but they have also been detected mixed with the crystalline silicate
dust in the nebulae around LBVs AG Car and R71 (\cite[Voors et al. 1999, 2000]{Voors_etal99}).  We
return to these features at the end of this paper.

The HDE316285 nebular spectra also exhibit pure rotational lines of H$_2$ arising from 
optically thin quadrupole transitions, from S(0) at 28.3 $\mu$m
through S(6) at 6.1 $\mu$m, though the S(4) 8.0 $\mu$m and S(6) lines are blended with PAH emission.   
{\it{This is the first known detection of rotational transitions of H$_2$ in the nebula of a hot 
massive star.}}  
\cite[St. Louis et al. (1998)]{StLouis_etal98} have previously detected the ro-vibrational 
1-0 S(1) 2.112 $\mu$m line in the NGC2359 nebula around WR7, in a region where the nebula may 
be interacting with a surrounding bow shock.  From this single line, however, St. Louis et al. 
could not deduce the excitation mechanism.  This is relavent to the debate on the contribution of
hot stars to the H$_2$ luminosity in starburst galaxies and ULIRGs, where flourescent excitation 
in the UV radiation fields of OB and WR stars has been generally preferred over collisionally-induced
emission for lack of cases of the latter in local settings.

The H$_2$ rotational lines are readily thermalized at moderate volume densities, 
allowing us to estimate H$_2$ column densities and excitation temperatures under 
the assumption of LTE in order to gain insight into the excitation conditions of the
line forming regions.  We follow excitation diagram methods developed by
\cite[Burton (1992)]{Burton92} and \cite[Gredel (1994)]{Gredel94}, and 
applied to comparable environments (as in, e.g., NGC7129 by 
\cite[Morris et al. 2004]{Morris_etal04} and references therein). The column density for
transition $j$ is $N_j$ = 4$\pi F_j / \Omega E_j g_j A_j$, where $E_j$ is the
energy of the upper level, $g_j$ is the statistical weight of ortho- and
para-transitions, and $A_j$ is the transition probability. The
line column density is related to the total column density and excitation
temperature  $T_{\rm{ex}}$ by $\ln(N_j/g_j) = ln(N_{\rm{tot}}/Q)-E_j/(kT_{\rm{ex}})$,
and may be solved for with a Boltzmann excitation diagram.  
The partition function $Q$ was determined with an ortho-para ratio of 3 under
the assumption of thermal equilibrium.  For the observed line fluxes per unit solid 
angle $F_j/\Omega$, we we have extracted over regions of 40~arcsec$^2$ and 62~arcsec$^2$ in the
5.3 -- 14 $\mu$m and 14 -- 35 $\mu$m ranges, respectively.  In these areas we
must view our measurements as spatially integrated.  The measured line intensities
were corrected for reddening $A_V$ = 6.0 mag (\cite[Hillier et al. 1998]{Hillier_etal98}). 
Measurement uncertainties of the S(3) -- S(5) lines arise 
predominantly from the blending with the PAH features and the flux calibration. 
The S(6)/S(5) or S(4)/S(5) line ratios could be also affected by an ortho-para
ratio different than the canonical (thermal equilibrium) value of 3, as in the 
presence of shocks, or by the superposition of 
multiple layers along the line of sight with different physical conditions. 
From our computed values of $\ln(N_j/g_j)$ and $E_{\rm{up}}$/K we
obtain estimates of the total H$_2$ column density $N_{\rm{tot}} \simeq
1.5 \times 10^{20} {\rm{cm}}^{-2}$ and excitation temperature $T_{\rm{ex}} \simeq
240$ K, averaged over the two positions. 

The relatively low value of $T_{\rm{ex}}$ at the high H$_2$ column density 
(compared to typical photodissociation regions) seem
quite reasonable for line emission in the outflow arising from collisional
(dissociative J-shock) excitation.  We favor the J-shock mechanism over
C-shocks in the outflow, based on the S(0) line intensity and strong 
[Si~{\sc{ii}}] 34.815 $\mu$m emission.  From J-shock models 
(\cite[Hollanbach \& McKee 1989]{HollanbackMcKee89}) we see also by the
presence of [Ne~{\sc{ii}}] 12.81 $\mu$m that the velocity is probably greater than 
50 km sec$^{-1}$.  

{\underline{\it{Is HDE316285 a hyper-luminous counterpart of the Pistol Star?}}}  In order 
to estimate the physical size
and age of the nebula around HDE316285 we must know the distance
to the star.  \cite[Hillier et al. (1998)]{Hillier_etal98} adopted 2 kpc, as a rough average
of literature values between 1 and 3.4 kpc, and conceded that there is no real constraint
on the distance.  At that time it was not recognized that HDE316285 lies in close (apparent)
proximity to the infrared SFR Sgr~D (see Fig. \ref{fig3}).  Assuming HDE316285 to be equidistant
with Sgr~D has the appeal of placing this B supergiant in the same dynamic ecosystem as the 
SFR and neighboring YSOs and maser sources, as well as a supernova remnant SNR~1.4-0.1 a 
few arcmin to the SE.  \cite[Mehringer et al. (1998)]{Mehringer_etal98} have estimated the 
distance of Sgr~D as 8 kpc based on H$_2$CO and CS line kinematics measured at the VLA.  At this
distance, the stellar luminosity of HDE316285 must be corrected from log($L_\star/L_\odot$) = 5.5
to 6.7$!$  This takes into account a small reduction to $T_{\rm{eff}}$ ($\simeq$14kK) in the 
most recent line-blanketed models, but does very little to moderate {\it{a stellar luminosity
matched only by the Pistol Star in the GC and $\eta$ Carina}}.  
Assuming a distance of 8 kpc, then the diameter of the nebula is 8.6 pc, and must have a 
dynamical age of at least 9300 years using an upper constraint of 450 km/s for 
the time-averaged expansion velocity set by the unresolved IRS lines ($\lambda/\Delta\lambda 
\simeq 650$).

We cannot settle comfortably with this new distance to HDE316285, unfortunately. 
\cite[Blum \& Damineli (1999)]{BlumDamineli99} have presented K-band imaging of Sgr~D, 
from which they argue that the SFR is located between 4-7.9 kpc (for $d_{GC}$ = 8 kpc) and
suggest $A_K$ = 1.9 mag from the nebular Br-$\gamma$/radio flux ratio.  No point sources
are detected in the K-band image within the core of Sgr~D, but focusing 
on sources within a few arcmin of the SFR, 18 of 34 sources have $A_K \leq$ 1.5 mag 
(versus $A_K$ = 3 for the GC itself). They devised a model with a uniform 
$A_K$ =  2.2 mag screen of extinction at 4 kpc from the Sun, matching 
the actual star counts.  HDE316285  has $A_V$ = 8 mag ($A_K \sim$ 0.8 mag), which is 
substantially lower, suggesting that the star is in the foreground. Indeed, University College London 
Echelle Spectrograph data taken in May 1995 show NaI D velocities of around +5 km/s, in agreement 
with a low distance to HDE316285 from comparison with the \cite[Brand \& Blitz (1993)]{BrandBlitz93} 
rotation curve using $d_{GC}$ = 8kpc. 

\begin{figure}[t]
\begin{center}
 \includegraphics[width=2.8in]{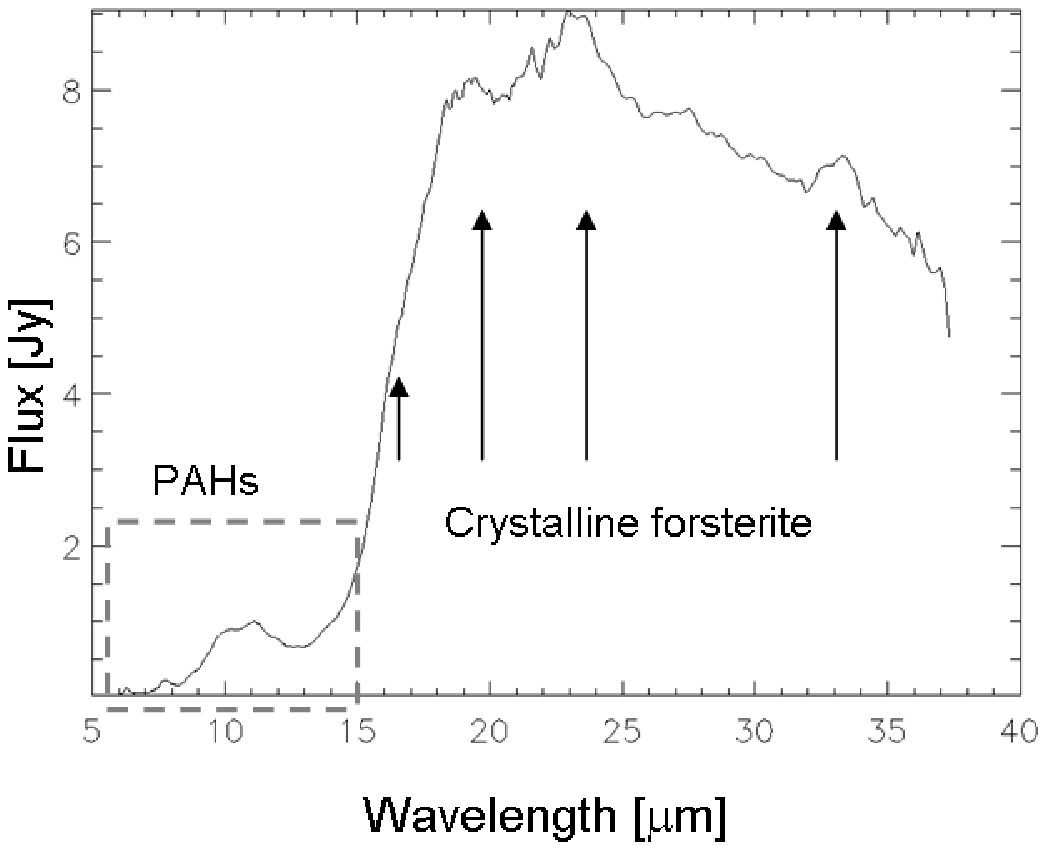}\includegraphics[width=2.7in]{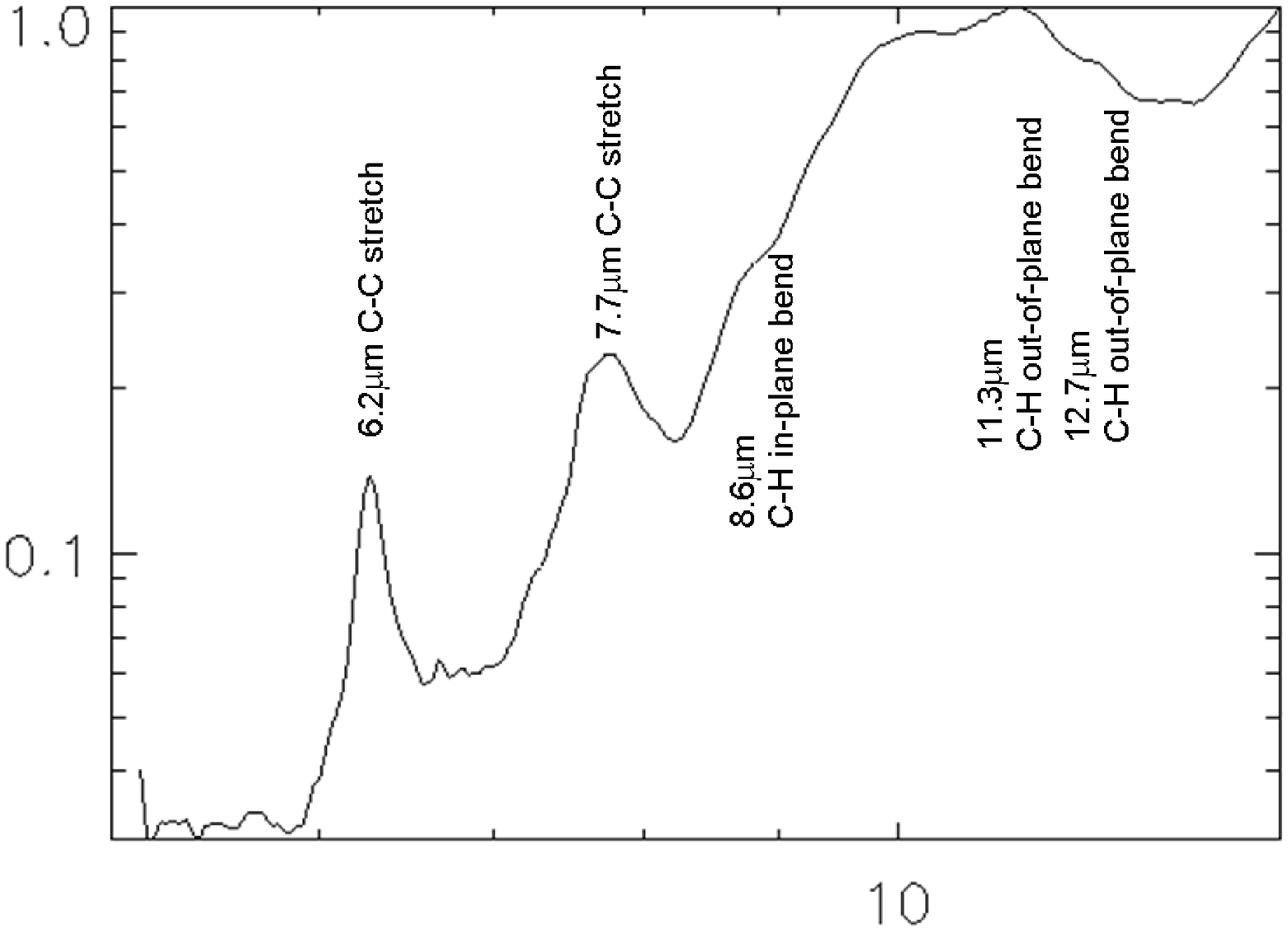}  
 \caption{{\em{Left:}} {\em{Spitzer}}/IRS spectrum of the LMC LBV R71, with principle crystalline
silicate bands and PAH features indicated.  Nebular dust properties have been studied by 
\cite[Voors et al. (1999)]{Voors_etal99} using ISO/SWS data to 25 $\mu$m.  {\em{Right:}} Zoom on 
the PAH bands in the IRS spectrum.  Was the material comprised of crystalline dust and hydrocarbons condensed in the C-rich envelope of an RSG phase? }
   \label{fig4}
\end{center}
\end{figure}

In summary, we are confronted with conflicting evidence on the distance to Sgr~D and HDE316285:
on the one hand H$_2$CO and CS radio line kinmatics place Sgr~D at 8 kpc, while near-IR modeling
of the reddened star population close to SFR can satisfy observed star counts at half the distance.  
Admittedly the latter method involves many more approximations and ensuing uncertainties, 
but the optical line (Na D) kinematics and comparatively low reddening of HDE316285 taken at 
their face values do not favor a large distance. If the B supergiant star is in the foreground, 
between 2 and 4 kpc, then either a substantial amount of foreground material is present also 
in the immediate surrounding ISM to interact with the stellar outflow, or else the 
H$_2$ lines are being formed in collisions between the present-day stellar wind and the 
outflow detected in our 24$\mu$m observations.

{\underline{\it{Implications and Outlook.}}}
The fact that HDE316285 is surrounded by a bipolar nebula supports the notion suggested
by \cite[Hillier et al. (1998)]{Hillier_etal98} that
this luminous, strong-winded B supergiant has recently passed through an LBV phase.  This 
is more plausible at a lower distance to the star with a commensurately lower 
limit age of the nebula, (i.e., $\geq 2300$).  We have also noted the lack of 
specific dust bands that are the signatures 
of crystalline material which normally condenses in the thick, slowly-expanding envelopes of 
RSGs, suggesting that the material forming the present-day nebula was rapidly ejected and 
could not crystallize under conditions of rapid cooling and low monomer densities.  
The explosive outflows from $\eta$ Carina also lack crystalline silicates, though there 
are a number of dust bands in this N-rich, C-depleted environment which are yet to be properly 
identified  (\cite[Morris et al. 1999]{Morris_etal99}).  In contrast, ISO spectroscopy of the 
LBVs AG~Car, WRA~751, and R71 (\cite[Voors et al. 1999, 2000]{Voors_etal00}) reveal 
crystalline properties of the O-rich dust and the presence of C-rich hydrocarbons. 
See Fig. \ref{fig4} for the IRS spectrum of R71.  
These properties are more consistent with formation during a RSG phase, supplying 
appropriate temperature and density condensation conditions as well as surface chemistries 
in which N(C)/N(O) $>$ 1, and fits with the rather low outflow velocities and time-averaged 
mass-loss rates which yield the nebular masses estimated by Voors et al.

We must be careful to consider certain potential attenuating effects on dust crystallinity
when interpreting the mass-loss history. Specifically,
silicates initially condensed amorphously in an eruption might be annealed under 
electron bombardment (\cite[Carrez et al. 2002a]{Carrez_etal02}).  Forsterite, an Mg-rich 
crystalline silicate detected in the aforementioned LBVs (see Fig. \ref{fig4}), can 
be formed at fluences of $\sim10^{17}$ e$^-$/cm$^2$ where (in the laboratory) 
particles are accelerated to several 10$^6$ eV.  This may occur in YSOs with magnetic 
fields.  OB supergiants are not magnetically active (except in the rare Ofp?e cases) and 
therefore the mechanism of electron bombardment would require acceleration by other means.
Conversely, amorphization of initially crystalline dust may occur under 
heavy proton bombardment, which may explain the lack of crystalline silicates in the ISM
(\cite[Carrez et al. 2002b]{Carrez_etal02b}, \cite[Brucato et al. 2004]{Brucato_etal04}).  
Both effects of electron (re-)annealing and proton amorphization require study applied to
BSG/RSG environments in order to improve the use of dust grain properties as 
tracers of the evolutionary state of the underlying star during condensation.

\begin{discussion}

\end{discussion}

\end{document}